\documentclass[prl,aps,twocolumn,floatfix,showpacs,superscriptaddress]{revtex4}
\usepackage{graphicx}
\usepackage{amsmath}
\usepackage[usenames]{color}

\begin{document}

\title{Adiabatic quantum pumping at the Josephson frequency}

\author{S. Russo}
\affiliation{Kavli Institute of Nanoscience, Faculty of Applied
Science, Delft University of Technology, Lorentzweg 1, 2628 CJ
Delft, The Netherlands}

\author{J. Tobiska}
\affiliation{NTT Basic Research Laboratories, Atsugi, Japan}

\author{T.M. Klapwijk}
\affiliation{Kavli Institute of Nanoscience, Faculty of Applied
Science, Delft University of Technology, Lorentzweg 1, 2628 CJ
Delft, The Netherlands}

\author{A.F. Morpurgo}
\affiliation{Kavli Institute of Nanoscience, Faculty of Applied
Science, Delft University of Technology, Lorentzweg 1, 2628 CJ
Delft, The Netherlands}

\date{\today}


\begin{abstract}
We analyze theoretically adiabatic quantum pumping through a
normal conductor that couples the normal regions of two
superconductor/normal metal/superconductor Josephson junctions. By
using the phases of the superconducting order parameter in the
superconducting contacts as pumping parameters, we demonstrate
that a non zero pumped charge can flow through the device. The
device exploits the evolution of the superconducting phases due to
the ac Josephson effect, and can therefore be operated at very
high frequency, resulting in a pumped current as large as a few
nanoAmperes. The experimental relevance of our calculations is
discussed.
\end{abstract}

\maketitle

\newpage

In a mesoscopic conductor in which electrons move phase
coherently, a direct current can flow  in response to a slowly
varying periodic perturbation and in the absence of any applied
bias. This phenomenon, known as quantum pumping, was first noticed
by Thouless\cite{Thouless}, who analyzed theoretically the
response of an electron system to a "traveling" periodic potential
$U(x-vt)$. The occurrence of quantum pumping requires that the
periodic perturbation consists of at least two independent
oscillating parameters $X_1(t)$ and $X_2(t)$, and that the
trajectory representing the perturbation in the parameter space
$(X_1,X_2)$ encloses a finite area\cite{Zhou,Brouwer2}. Indeed,
the proposal of Thouless satisfies these requirements since even
the simplest travelling periodic potential $U(x-vt)=U_0
\sin(x-vt)$ can be written as $U(x-vt)=X_{1}(t) \sin(\frac{2 \pi
x}{\lambda})+X_{2}(t) \cos(\frac{2 \pi x}{\lambda})$, with
$X_{1,2}(t) =X_{1,2} \cos(\frac{2 \pi t}{\tau}+ \phi_{1,2})$. When
the cyclic perturbation is slower than the electron dwell time in
the conductor, adiabatic pumping occurs and the system remains in
thermodynamic equilibrium. In this case, the pumped charge can be
expressed as a function of the scattering matrix and of its
derivatives with respect to the pumping parameters $X_1$
and $X_2$ \cite{Brouwer2}.\\
Attempts to investigate experimentally adiabatic quantum pumping
have been made using electrostatically defined quantum dots in
GaAs-based heterostructures \cite{Switkes}. In such a system,
pumping is induced by oscillating voltages applied to the gate
electrodes defining the dot. Although signatures of pumping
signals may have been observed, the experiments are hindered by
rectification effects originating from parasitic coupling of the
$ac$ signal applied to the gates \cite{Brouwer3}. Many other
proposals of devices have been put foward in the literature, in
which different physical quantities have been used as pumping
parameters such as a time-varying magnetic field, the height of a
tunnel barrier, etc.\cite{Vinokur,Entin-Wohlman,Ch6Wang}. Often,
however, these proposals do not consider the difficulties
involved in the experimental realization.\\
\begin{figure}[h]
    \centering
   \includegraphics[width=\columnwidth]{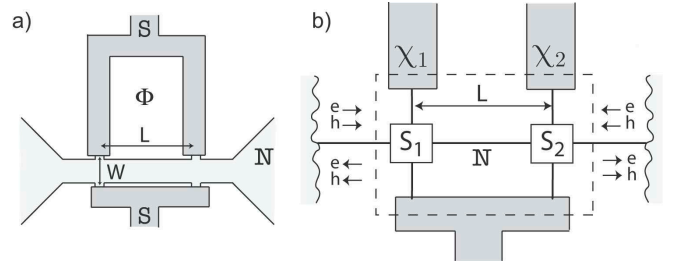}
   \caption{a) Layout of the proposed quantum pump, consisting of two SNS Josephson
   junctions in a SQUID geometry with a common normal region
   (The dark grey areas represent the superconductor electrodes and light grey the normal conductor). b)
   Diagramatic representation of the scattering matrix
   of the device.}
        \label{Pumping_fig1}
\end{figure}
Here we demonstrate theoretically the occurrence of pumping in a
system of electrons and holes in a metallic conductor coupled to
superconductors, where the pumping parameters are the phases of
the superconducting order parameters in two different
superconducting contacts. This system can operate at very high
frequency without the need of feeding microwave radiation, simply
by exploiting the evolution of the superconducting phases due to
the \textit{ac} Josephson effect. As a consequence, measurable
pumped currents as large as a few nanoAmperes, can be expected,
while avoiding spurious effects that affected previous
experiments.\\
Fig. \ref{Pumping_fig1}a shows a schematic representation of the
circuit that we propose. Two superconducting/normal
metal/superconducting (SNS) Josephson junctions are connected in
parallel via a superconducting ring, and their N regions are
additionally coupled by a normal metal bridge. Andreev reflection
\cite{Ch6Andreev} of electrons and holes takes place at each NS
interface, resulting in a phase shift of the particle wavefunction
which at the Fermi energy is given by $\pm \chi_{1,2}$, the phase
of the superconductor order parameter at the two different
superconducting contacts (the sign - is for reflection from hole
to electron; the sign + for the reverse process). Hence, the total
scattering matrix ($S_{tot}$) of the normal metal bridge
connecting the left and right reservoirs depends on the quantities
$X_{1} = e^{i \chi_{1}}$ and $X_{2} = e^{i\chi_{2}}$. We want to
see if a direct current can flow in the normal metal bridge when
$X_{1}$ and $X_{2}$ are used as pumping
parameters.\\
The appealing aspect of such a device is the way in which the
pumping parameters can be driven at high frequency, and their
relative phase controlled. Specifically, the pumping parameters
become time-dependent when a constant voltage $V_{dc}$ is present
across the SNS junctions (e.g., by biasing the junctions with a
current higher than their critical current), since then $X_{1}(t)$
and $X_{2}(t)\propto e^{i \frac{2eV_{dc}}{\hbar}t}$ owing to the
$ac$ Josephson effect \cite{Likharev}. Similarly to what happens
in superconducting quantum interference devices (SQUIDS), the
phase difference $\varphi$ between $X_{1}$ and $X_{2}$ can be
easily controlled by applying a magnetic flux $\Phi$ to the
superconducting loop, so that $\chi_{2}=\chi_{1}+\varphi$, with
$\varphi=2\pi \Phi/\Phi_0$
($\Phi_0=h/2e$).\\
\begin{figure}[h]
    \centering
    \includegraphics[width=\columnwidth]{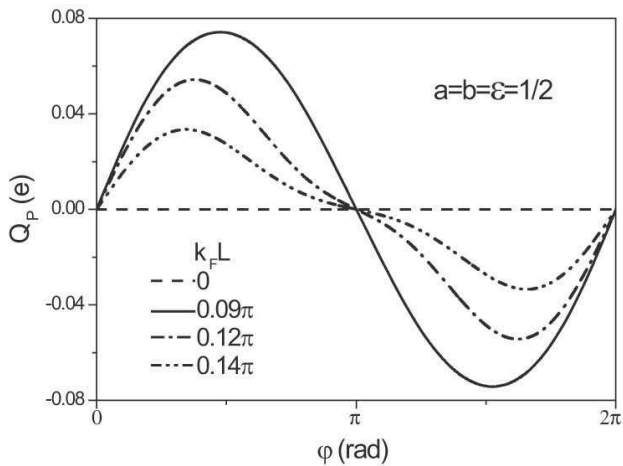}
    \caption{Pumped charge per cycle $Q_P$ as a function of phase difference $\varphi$
    between the pumping parameters, for the case $a=b=\varepsilon=1/2$
    and different values of $k_F L$.}
        \label{Pumping_fig2}
\end{figure}
To demonstrate the occurrence of pumping, we model the system in
the simplest possible way. We confine ourselves to the case of a
fully phase coherent system at $T=0$ K. The normal conductor is
taken to consist of one channel supporting ballistic motion, and
the separation between the Josephson junctions $L$. The N/S
interfaces are all supposed to be perfectly transparent,
\textit{i.e.} the probability of Andreev reflection is
unity.\\
The pumped current is equal to the charge pumped per cycle,
multiplied by the pumping frequency. The calculation of the charge
pumped per cycle follows the approach developed by
Brouwer\cite{Brouwer2}, modified to take into account the presence
of the superconducting electrodes\cite{Ch6Wang}. The relation
between the charge $Q_{P,m}$ pumped in one of the two reservoirs
(labeled by $m=1,2$) and the
scattering matrix reads:\\
\begin{equation}
  Q_{P,m}=e \int_0 ^{\tau} dt \left( \frac{dn_m}{dX_1}\frac{dX_1}{dt}+
  \frac{dn_m}{dX_2}\frac{dX_2}{dt} \right), \label{Brouwer_formula1}
\end{equation}
in which:\\
\begin{equation}
\frac{dn_m}{dX_{1,2}}=\frac{1}{2 \pi} \sum_{i,j}\gamma_{ij}
\mbox{Im}\frac{\partial (S_{tot})_{ij}}{\partial
X_{1,2}}(S_{tot})_{ij}^*, \label{Brouwer_formula2}
\end{equation}
where $\tau$ is the period of one pumping cycle
($\tau=\frac{2\pi}{\omega_J}$, with $\omega_J=\frac{2e
V_{dc}}{\hbar}$). In Eq. \ref{Brouwer_formula2}, the sum over $i$
extends to the electron and hole ($e,h$) channels in both leads.
The sum over $j$ is performed over the electron and hole channels
only in the lead connected to the reservoir \textit{m} for which
the pumped charge is calculated. The function $\gamma_{ij}$ is
equal to +1 when the element $S_{i,j}$ of the scattering matrix
corresponds to a process in which a current is pumped
\textit{from} lead \textit{m}, $\gamma_{ij}=-1$ when a current is
pumped \textit{into} lead \textit{m}. This difference in sign is
due to the fact that electrons and holes contribute oppositely to
the pumped charge. Since the electron and hole contributions to
the pumped charge could exactly compensate each other, it is not
obvious \textit{ a priori}
whether a net charge can be pumped.\\
The problem of computing the pumped charge is then reduced to the
calculation of the total scattering matrix $S_{tot}$ of electrons
and holes in the normal conductor bridge (see Fig.
\ref{Pumping_fig1}b), as a function of the parameters $X_1$ and
$X_{2}$. The calculation is lengthy but conceptually
straightforward (calculations were done using
\textit{Mathematica}$\texttrademark$). We consider a perfectly
symmetric configuration with two identical SNS junctions, which
are also identically coupled to the normal metal bridge. For each
junction, the coupling is described by a "beam-splitter$\lq\lq$
\cite{Buttiker}, whose scattering matrix ($S_{1,2}$) is assumed to
be energy independent (\textit{i.e.}, it is the same for electrons
and holes). We have chosen the simplest expression compatible with
unitarity
and time reversal symmetry. The expression reads:\\
\begin{equation}
S_{1,2}=\begin{pmatrix}
  a  & \sqrt{\frac{\varepsilon}{2}} & b & \sqrt{\frac{\varepsilon}{2}} \\
  \sqrt{\frac{\varepsilon}{2}} & -a  & \sqrt{\frac{\varepsilon}{2}} & -b \\
  b & \sqrt{\frac{\varepsilon}{2}} & a  & \sqrt{\frac{\varepsilon}{2}} \\
  \sqrt{\frac{\varepsilon}{2}} & -b & \sqrt{\frac{\varepsilon}{2}} &
  -a
\end{pmatrix},
\end{equation}
where $\epsilon$ varies between 0 and $1/2$ ($\epsilon/2$ is the
probability for an incoming particle to be deflected towards one
of the superconductors). The amplitudes for backscattering
\textit{a} and direct transmission \textit{b} across the beam
splitter satisfy the relations $a^2+b^2+\varepsilon=1$ and
$\varepsilon/2ab=1$, imposed by unitarity. For every fixed value
of $\varepsilon$ two solutions, with $a>b$ and $b>a$, are possible
and we considered both cases (for $\varepsilon=1/2$ the two
solutions coincide and $a=b=1/2$).\\
Mixing of electrons and holes only occurs at the interface with
the superconductors. Having assumed perfect transparency at the NS
interfaces, the matrix describing Andreev reflection in the
"vertical" branches of the circuit (see Fig.\ref{Pumping_fig1})
depends only on the phase $\chi$ of the superconducting order
parameter. It reads:\\
\begin{equation}
S_{\mathrm{AR}}=\begin{pmatrix}
  0 & r_{\mathrm{he}} \\
  r_{\mathrm{eh}} & 0
\end{pmatrix}
=
\begin{pmatrix}
  0 & -ie^{i \chi} \\
  -ie^{-i \chi} & 0
\end{pmatrix}.
\end{equation}
To calculate the total scattering matrix of the device we first
calculate the scattering matrix associated to transport across
only one SNS junction. We then consider the two SNS junction
connected in series, \textit{i.e.} we consider all the multiple
reflection processes in the normal metal bridge, taking into
account the corresponding dynamical phases acquired by electrons
and holes. The result is the scattering matrix $S_{tot}(X_1,X_2)$
that mixes the electron and hole channels in reflection and
transmission.\\
\begin{figure}[h]
    \centering
   \includegraphics[width=\columnwidth]{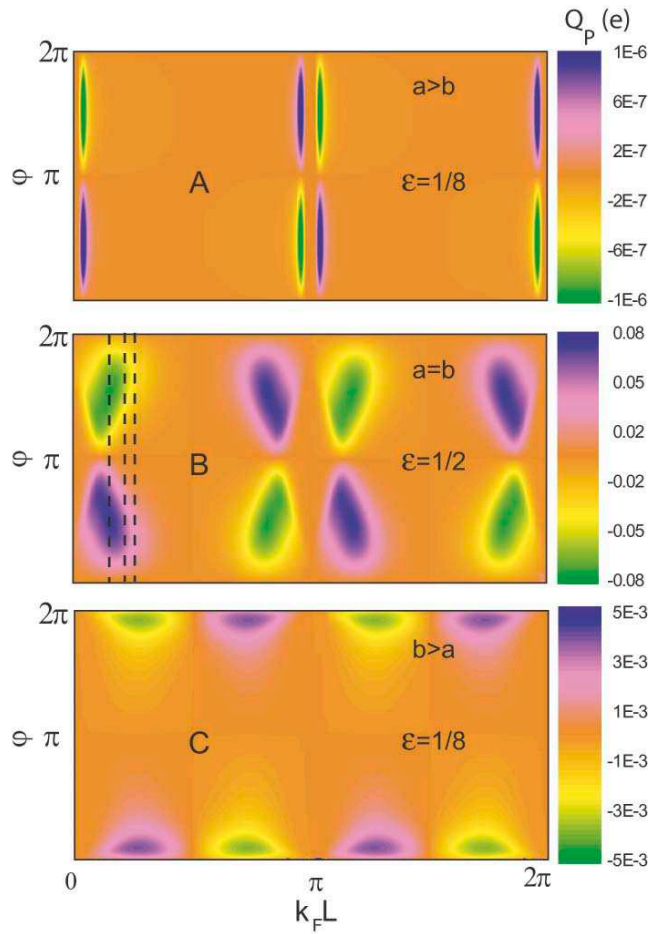}
   \caption{Color scale plots of $Q_P$ as a function of $k_F L$
   and
   $\varphi$ for three distinct cases of amplitude scattering on the beam splitter:
   $\varepsilon=1/8$ for $a>b$ (top),
   $a=b=\varepsilon=1/2$ (center) and $\varepsilon=1/8$ for $b>a$ (bottom).
   The dashed lines in the center panel
   correspond to the curves shown in Fig. 2.}
       \label{Pumping_fig3copy2}
\end{figure}
Having determined $S_{tot}$ we obtain the pumped charge from Eqs.
\ref{Brouwer_formula1} and \ref{Brouwer_formula2}. As shown in
Fig. \ref{Pumping_fig2} we find that, unless $k_F L = n \pi$ (with
$n$ integer), the pumped charge is a non-zero, anti-symmetric, and
$2 \pi$-periodic function of $\varphi$, as expected. The $2 \pi$
periodicity in conjunction with the antisymmetry imply that the
pumped charge has to vanish when $\varphi = \pm \pi$. This is the
case since for $\varphi=\pi$ the trajectory in the space of the
pumping parameters $(X_1,X_2)$ does not enclose a finite area. In
addition, the antisymmetry of $Q_P$ with respect to $\varphi$ also
implies that the sign of the pumped current changes when reversing
the sign of the relative phase of the two superconducting
junctions. This results in the antisymmetry of the pumped current
versus applied magnetic flux $\Phi$ (that determines the phase
$\varphi$), and provides a distinctive feature that should
facilitate the experimental identification of
the phenomenon.\\
Fig. \ref{Pumping_fig3copy2}A, B, and C summarize the outcome of
our calculations for the different cases $a>b$, $a=b$, and $a<b$.
We first discuss the features of the results that are common to
all three cases. We always find that the pumped charge does not
depend on the separation $W$ between the beam splitters and the
superconducting leads (see Fig. \ref{Pumping_fig3copy2}). This is
due to the phase conjugation \cite{Takayanagi} of electrons and
holes at the Fermi energy, since the dynamical phase acquired by
an electron propagating from the beam splitter to the
superconducting interface is exactly compensated by the phase
acquired by the Andreev reflected hole. In all cases the
dependence of $Q_P$ on $L$ is periodic for all values of $\varphi$
and $\varepsilon$, with period given by $k_FL=\pi$ ($k_F$ is the
Fermi wave vector). This implies that the pumped charge is
sensitive to the geometry of the device on the scale of the Fermi
wavelength $\lambda_F$, indicating that charge pumping in the
device considered here is a sample specific phenomenon. That this
should be so is not obvious a priori : owing to phase conjugation,
one may have expected the pumped charge to show a component
independent of the precise geometry of the device
\cite{note1}.\\
The magnitude of the calculated pumped charge strongly depends on
$\varepsilon$. The maximum pumped charge is approximately 0.1
electron per cycle, for $\varepsilon=1/2$ ($a=b$). For small
$\varepsilon$, the magnitude of $Q_P$ decreases with decreasing
$\varepsilon$ (and eventually vanishes for $\varepsilon=0$) both
when $a>b$ and $b>a$. The dependence of $Q_P$ on $\varphi$ and
$k_F L$, however, is different in the two
cases.\\
When $a>b$ and for small values of $\varepsilon$ (\textit{i.e.}
$a\sim 1$) the bridge connecting the two SNS junctions is only
weakly coupled to the reservoirs, because backscattering at the
beam-splitters is the dominant process (see Fig.
\ref{Pumping_fig3copy2}A). In this regime, sharply defined
resonances appear in the conductance of the system when
$k_FL=n\pi$ (with $n$ integer), due to the presence of quasi-bound
states in the bridge connecting the two SNS junctions. When
$k_FL=n\pi$ the energy of a quasi-bound state aligns with the
Fermi levels in the reservoirs. Interestingly, the pumped charge
is also significantly different from zero only when $k_FL$ is
close to being a multiple of $\pi$. This suggests a close link
between pumping and the presence of resonances due to quasi-bound
states in the system, as already noted by others in different
contexts\cite{Entin-Wohlman}. This link is further supported by
observing that increasing $\varepsilon$ from 0 to $1/2$
-corresponding to increasing the broadening of the quasi-bound
states- results in a broader range of values of $L$
for which charge pumping is observed (Fig. \ref{Pumping_fig3copy2}B).\\
In the case $b>a$, the behavior of the pumped charge for small
values of $\varepsilon$ is qualitatively different (see Fig.
\ref{Pumping_fig3copy2}C). In this regime, the dominant process at
the beam splitters is direct transmission. Therefore electrons and
holes have only a small probability to be deflected from the
normal bridge to the N/S interfaces. However, if they are
deflected, they perform many Andreev reflections in one of the SNS
junctions before they can escape again to the normal metal bridge.
As a consequence, along the dominant trajectories responsible for
pumping, electrons and holes have a large probability to acquire a
phase $e^{iN\chi}$ (with different, and even large, integer values
of $N$), rather than simply $e^{i\chi}$. This causes the phase
dependence of the pumping signal to be richer in harmonics and,
consequently, to exhibit very strong deviations from a simple sine
dependence, as seen from Fig. \ref{Pumping_fig3copy2}C.\\
Having established the occurrence of adiabatic quantum pumping, we
briefly discuss some of the advantages of the proposed device. The
use of the $ac$ Josephson effect to generate the time dependence
of the pumping parameters (the superconducting phases in our case)
should allow operation at frequencies of several hundreds GHz. In
fact, with superconductors such as Nb, NbN, or NbTiN, values for
the superconducting gap $\Delta$ corresponding to frequencies in
excess of 1 THz are possible, so that our superconducting pump can
operate at a few hundreds GHz when the voltages applied across the
SNS junctions is still sufficiently lower than $\Delta$. At a
Josephson frequency of 100 GHz, the pumped current can exceed 1
nA, which is easily measurable. Note that, since the pumping
parameters are coupled to the electron-hole wave functions via
Andreev reflection, the coupling will remain good at these high
frequencies. In addition, the fact that no external microwave
signals need to be fed into the circuit to drive the pumping
parameters implies that only a negligible high-frequency power
will be irradiated, thereby minimizing the possibility of
rectification effects known to cause problems in other
systems\cite{Brouwer3}. For the practical realization of the
proposed superconducting pump we suggest the use of a ballistic
InAs-based two-dimensional electron gas as normal conductor.
Present technology enables the reduction of the number of
conducting channels to $\approx 10$ \cite{Alberto}, which is
important since the predicted effect is of the order of one
channel. The use of InAs also enables the realization of the
needed highly transparent contacts to superconductors
\cite{Ch6Heida}. Furthermore, in ballistic devices in which the
distance between the two SNS junction is $L \simeq 1 \mu$m, the
typical propagation time in the device will be of the order of
$L/v_F \ 10^{-12}$ s ($v_F \simeq 10^6$ m/s is typically realized
in InAs heterostructures). This is ten times faster than the
period of an $ac$ pumping signal oscillating at 100 GHz, ensuring
that the dwell time of electrons is much shorter than the period
of the ac pumping signal, as it is needed for the device to
operate in the adiabatic regime.\\
We gratefully acknowledge M. Blauboer, Y.V. Nazarov, Y. Tokura and
P.W. Brouwer for useful discussions. This research is financially
supported by the Dutch Foundation for Fundamental Research (FOM)
and NWO (Vernieuwingsimpuls 2000).

\end{document}